# Rejoinder for the discussion of the paper "A novel algorithmic approach to Bayesian Logic Regression"

Aliaksandr Hubin[*], Geir Storvik[†] and Florian Frommlet[‡]

## 1 Introduction

We would like to begin this rejoinder with expressing our sincere gratitude to all of the discussants for their interesting and thought-provoking comments and remarks. We also feel heartily thankful to the editorial board of Bayesian Analysis for giving us the opportunity to publish our paper entitled "A novel algorithmic approach to Bayesian logic regression" (Hubin et al., 2020a) as a discussion article. Logic regression is a tool to model non-linear relationships between binary covariates and some response variable by constructing predictors as Boolean combinations. The number of possible logic expressions grows exponentially with the number of binary variables involved, making the model search significantly harder with the increasing complexity of Boolean combinations. Due to Boolean equivalence, it is in fact almost impossible to specify the full model space a priori even for a relatively small number of covariates.

Our primary goal is to identify those logic expressions which are associated with the response variable. To this end, we want to estimate posterior probabilities of logic expressions within the framework of generalized linear models. The major contributions of our paper are two-fold: Firstly, we have introduced novel model priors for Bayesian logic regression (BLR), which yield good power to detect important logic expressions while controlling the number of false positive discoveries. Secondly, we have introduced a novel genetically modified mode jumping Markov chain Monte Carlo (GMJMCMC) algorithm to efficiently explore the space of logic regression models.

The main idea of GMJMCMC is to embed the mode jumping Markov chain Monte Carlo (MJMCMC) algorithm (Hubin and Storvik, 2018) into the iterative setting of a genetic algorithm. Populations for the genetic algorithm consist of relatively small sets of logic expressions. Any such subset forms a well defined model space which allows to run MJMCMC. The population is then regularly updated in such a way that the algorithm is guaranteed to be irreducible in the model space of all logic regression models. This is required for asymptotic unbiasedness of the estimated posterior probabilities, as we will discuss in more detail below. Although GMJMCMC is not a proper MCMC algorithm (in the sense that its stationary distribution does not coincide with the target distribution of interest), renormalized estimates of the posterior probabilities are readily available.

---

[*]Norwegian Computing Center, aliaksandr.hubin@nr.no
[†]Department of mathematics, University of Oslo, geirs@math.uio.no
[‡]Department of Medical Statistics (CEMSIIS), Medical University of Vienna, florian.frommlet@meduniwien.ac.at

The discussants have pointed out several interesting extensions and open problems. We have structured the rejoinder according to different topics while trying to address all the points raised by the discussants. We also provide several interesting extensions of the model. Finally, we give a brief tutorial on the relevant part of our R-package EMJM-CMC http://aliaksah.github.io/EMJMCMC2016/ dealing with BLR. This should facilitate the practical application of our new methodology developed in Hubin et al. (2020a).

## 2  Applications of Bayesian logic regression

We very much appreciate that Ruczinski et al. (2020) have pointed out important applications of logic regression outside of genetics. Our emphasis on genetic applications was not meant to indicate limitations of the usefulness of logic regression in other areas but rather reflects our own previous research interests. Also, the applications mentioned by Bogdan et al. (2020) in the context of multicolored graphical models sound quite interesting. We are however more sceptical whether logic regression, in whichever form, will ever be applicable directly to association studies of outbred populations, *where the number of genetic variants is much larger than in controlled populations*. For large numbers of binary covariates, already the number of pairwise logic expressions becomes prohibitively large to apply logic regression, both in terms of algorithmic feasibility and in terms of having sufficient power while controlling type I error. Realistic applications of logic regression (with the aim of identifying true logic expressions) will most likely be restricted to applications with a few hundred binary covariates unless technologic advances allow one day to efficiently resolve the $\mathcal{NP}$ hard combinatorial problem of model search. However, one might consider bagging and boosting to obtain scalable versions of logic regression for prediction.

In this rejoinder, we also discuss extensions of Bayesian logic regression, allowing for non-binary predictors and latent Gaussian variables to be included into the model. This could further extend the applications of Bayesian logic regression methodology to such fields as epidemiology, spatio-temporal statistics, environmetrics and econometrics. For example, in Hubin et al. (2020c), a model with latent Gaussian processes (where a subset of predictors are binary) was used for the analysis of DNA methylation. The paper discusses the potential of using logic expressions of the binary predictors as a direction for further research. With the extensions of BLR provided in this rejoinder, it would become feasible to perform logic regression in the settings of Hubin et al. (2020c).

Both Ruczinski et al. (2020) and Bogdan et al. (2020) commented on the lack of correlated regressors in our simulation studies. This was mainly due to the fact that for the sake of comparison we wanted to use the scenarios from Fritsch (2006). For the more complex scenarios, we simply extended these scenarios by adding logic expressions of a higher order. In our sensitivity analysis, though, we considered one scenario with correlations (but only for one *true* leaf), where reasonably good results were obtained when the correlation of a mis-specified leaf $r$ was being varied from 0.1 to 1. However, specifically to respond to the remarks from Bogdan et al. (2020), who hypothesised that our approach would only work under independence, we will provide some additional simulation results, where we consider regressors with different correlation structures.

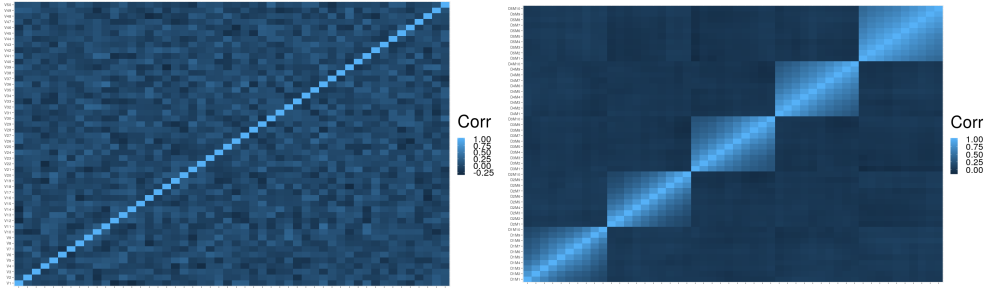

Figure 1: Correlation structure of the simulated covariates with a general correlation structure (left) and from QTL back-cross (right).

## 2.1 Simulation study with correlated regressors

In this study, we simulate the data using $p = 50$ regressors with two different types of correlation structure: The first one is rather general and uses fairly weak correlations, whereas the second one is typical for QTL mapping and gives very strong correlations. For the first scenario, we consider covariates which are marginally distributed according to $X_j \sim$ Bernoulli$(0.5), j \in \{1, ..., 50\}$. The correlation matrix is obtained using the approach from Joe (2006), which allows to generate positive definite matrices where all pairwise correlations are i.i.d. from a Beta distribution $B(a, a)$ linearly transformed to the interval (-1, 1). The parameter of the Beta distribution equals $a = alphad + (p-2)/2$, where $alphad > 0$ can be chosen. In our case for $p = 50$ and $alphad = 5/2$ it holds that the pairwise correlation lies between -0.2 and 0.2 with probability 0.85 and between -0.3 and 0.3 with probability 0.97. Correlations with an absolute value larger than 0.4 are extremely unlikely. Multivariate binary random variables $X_j, j \in \{1, ..., p\}$ with such correlation structures are then simulated by thresholding normal distributions as described by Leisch et al. (1998). A typical correlation structure of covariates generated by this approach is shown in the left panel of Figure 1.

The second scenario is based on the classical back-cross design for QTL mapping. We used the R/QTL package (Broman et al., 2003) to generate a map of 5 chromosomes of different lengths ranging from 100cM to 40cM with 10 equidistant markers per chromosome, see Figure 2. For experimental populations, there is a direct relationship between the genetic distance between markers on the same chromosome and their correlation as described in any textbook on QTL mapping (Chen, 2016). The corresponding correlation structure from simulated genotypes of $n = 1000$ individuals from a back-cross design is illustrated by the heatmap in the right panel of Figure 1. One can see that the correlations between markers on the same chromosome are very strong, getting close to 0.9 for neighbouring markers.

Response variables $Y$ are simulated from the data generating model of Scenario 5 from Hubin et al. (2020a), where $Y \sim N(\mu, 1)$, with

$$\mu = 1 + 1.5L_1 + 3.5L_2 + 9L_3 + 7L_4 \ . \tag{1}$$

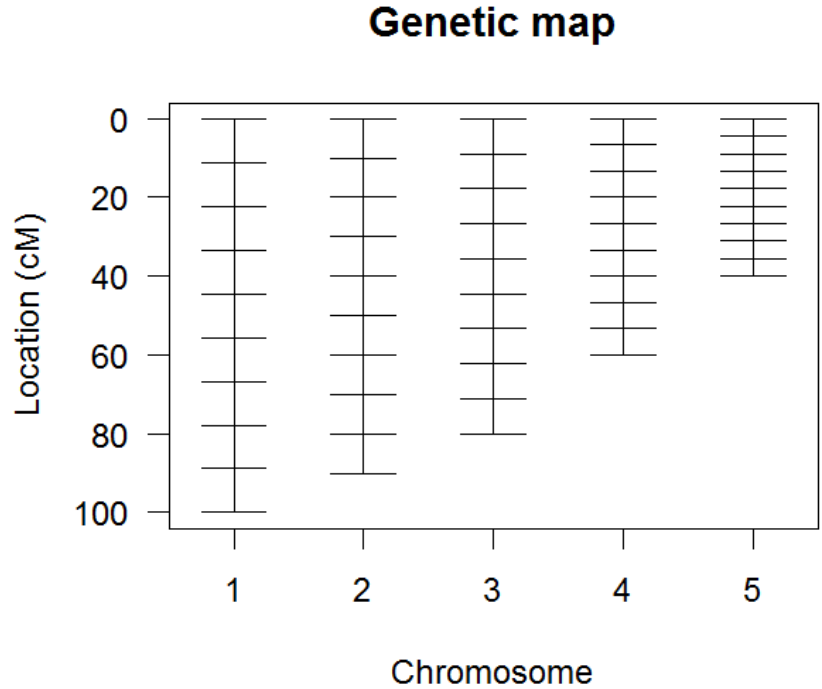

Figure 2: Genetic map of markers on five chromosomes of different length (given in centiMorgan). For the second scenario of our simulation study, these marker positions are used to simulate genotype data from a back-cross design. The closer markers on the same chromosome are lying the stronger will be the correlation of the corresponding genotype data.

The exact definition of the trees $L_1 - L_4$ is given in Table 1 below and is equivalent to the definition in Table 2 of our original article. For the QTL mapping scenario, the responses were generated for each simulation replicate after randomly permuting the order of the genetic markers. In this way, we considered different patterns of correlations between the leaves of the data generating model. For both correlation structures, we generated $N = 100$ datasets with $n = 1000$ observations. Every data set was analysed with the Jeffreys prior and with the robust g-prior using GMJMCMC with the same tuning parameters as in Scenario 5 of the original article.

We appreciate the comment of Schwender and Ickstadt (2020) that for the higher order interactions $L_3$ and $L_4$ the effect sizes are unrealistically large. However, as illustrated by our sensitivity analysis, if one wants to have sufficient power do detect more complex logic expressions with realistic effect sizes then one will need a much larger sample size. This would be potentially feasible for real data analysis (by means of simply collecting more observations) but not for a simulation study with hundreds of simulation runs. In any case, our goal here is to show that correlated regressors are not an impediment for our approach.

|  | General | | QTL | |
|---|---|---|---|---|
|  | **Jef.** | **R. g** | **Jef.** | **R. g** |
| $L_1 = X_{37}$ | 1.00 | 1.00 | 0.83 | 0.85 |
| $L_2 = X_2 \wedge X_9$ | 0.98 | 0.99 | 0.82 | 0.81 |
| $L_3 = X_7 \wedge X_{12} \wedge X_{20}$ | 0.96 | 0.99 | 0.92 | 0.92 |
| $L_4 = X_4 \wedge X_{10} \wedge X_{17} \wedge X_{30}$ | 0.66 | 0.66 | 0.20 | 0.24 |
| Overall Power | 0.90 | 0.91 | 0.69 | 0.71 |
| FP | 0.78 | 0.73 | 2.02 | 2.01 |
| FDR | 0.13 | 0.13 | 0.39 | 0.38 |
| WL | 9 | 6 | 108 | 98 |

Table 1: Results for the additional simulation scenarios with correlated binary covariates. Power for individual trees, overall power, the expected number of false positives (FP) and FDR are compared for GMJMCMC using either Jeffreys prior or the robust g-prior under the general correlation structure and the correlation structure from QTL mapping with back-cross design.

Table 1 summarizes the results of our simulations with correlated regressors. For the first scenario with the general structure, correlations are ranging between 0 and 0.5 in absolute values. Comparing the results with Table 2 from the original manuscript which was based on independent regressors the differences are relatively small. Only for $L_4$, there is a decrease in power, for Jeffreys prior from 0.89 to 0.66 and for the robust g-prior from 0.9 to 0.66. On the other hand, the number of false positives increases for Jeffreys prior from 37 to 78 and for the robust g-prior from 28 to 73. It has to be expected that the performance of logic regression becomes a little worse under correlation but GMJMCMC is still behaving very well for the scenario with a general correlation structure. Jeffreys prior and robust g-prior perform almost equally well with only a very slight advantage of the latter.

The results in Table 1 for our second correlation structure from QTL mapping are based on the strict definition that only discoveries of trees from the data-generating model itself are counted as true positives. While there is some loss of power, the results for the first three logic expressions are still quite satisfactory. Only for $L_4$, the estimated power becomes unacceptably low. At the same time, the number of false positives, as well as the number of wrongly detected leaves, increases substantially. For QTL mapping, the correlation between neighboring markers often is so strong, that it becomes extremely difficult to distinguish between them. For that reason, in simulation studies for QTL-mapping, one often takes the approach that the detection of a marker strongly correlated with a QTL is still counted as a true positive. If we take such an approach and consider markers within a range of 15cM as correct representatives of a leaf from the data generating model then we get slightly better results. In particular, the number of wrong leaves goes down from 108 to 50 for Jeffreys prior and from 98 to 58 for the robust g-prior. Extending the window for defining true positives would further reduce the number of wrongly detected leaves.

## 3 Prior related aspects

Clarte and Robert (2020) criticize several aspects of our choice of priors. We fully agree that the use of improper priors is debatable and should be done with great care. We had stated this *explicitly* already in the original article. From a theoretical point of view, our preference would be mixtures of g-priors. As a representative, the robust g-prior is implemented within our package. However, given the strong popularity of the BIC criterion, we wanted to study the performance of this choice as well. Our description of BIC as an approximation of the marginal likelihood under Jeffreys prior could indeed have included a discussion of its weak points. As Clarte and Robert (2020) remark, the good performance of the BIC choice is most likely connected with applying the Laplace approximation of the marginal likelihood. However, in the case of the Gaussian linear model, the approximation is exact.

The main empirical point, though, is that in all our examples from the original manuscript, the BIC measure as an approximation for the marginal density performed better than the analytical expression under the robust g-prior, both in terms of evaluation metrics and speed. Under the correlated designs provided in this rejoinder, the robust g-prior slightly outperforms Jeffreys prior in terms of evaluation metrics but the BIC choice still performs rather well. Moreover, the running time of GMJMCMC under Jeffreys prior (having all of the tuning parameters of the algorithm fixed) is still significantly shorter.

Note that the main contributions of our approach are: a) introducing novel model priors and b) a new search algorithm, whilst for the choice of the parameter priors and the calculation of the marginal densities we are using already established procedures. For example, our approach is fully compatible with integrated nested Laplace approximations (INLA) (Rue et al., 2009) and all of the parameter priors available there can be used. More generally, the R-package we have developed allows the users to easily specify their own method of calculating the marginal likelihood (whatever they prefer and/or what is available for their specific model: analytical integration, Monte Carlo based approximation, or other approximations) for their own choice of parameter priors. This flexibility allows extending the method easily to broader classes of logic regression models. In Section 5, for instance, we describe an extension to latent Gaussian models with both logic and non-logic covariates, where alternative types of parameter priors are possible and the marginal likelihood is computed via integrated nested Laplace approximations (INLA) (Rue et al., 2009).

Also note, that in both Bayarri et al. (2012) and Li and Clyde (2018), priors on models are indirectly obtained through priors on the regression parameters. In our approach, we include specific priors on model complexity as well. This is done via equations (3) and (4) in the main paper. The theoretical properties of combining model and parameter priors definitely require further distinguished research, which, we feel, lies slightly outside the scope of this rejoinder.

## 4 Algorithmic aspects

Given that one of the main contributions of this manuscript was the development of the GMJMCMC algorithm, it is no surprise that many comments of the discussants were concerned with the algorithm. We will start with replying to some questions which are simple to answer, then give a more detailed recap of the MJMCMC algorithm (Hubin and Storvik, 2018) and finally discuss some questions on the parameter settings of GMJMCMC.

Ruczinski et al. (2020) wondered whether covariates which are not logic can be easily combined with Boolean combinations in the model. The answer is *yes*. We will discuss this extension in Section 5.2 and provide an example in the tutorial in Section 6.3. Ruczinski et al. (2020) also suggested a simple two-stage approach where one first checks whether the covariates have any association with the response variable at all and one only then applies logic regression. This is, of course, a viable approach which can be easily adopted. In practice, this could save resources by avoiding to run computationally costly inference on BLR.

Whilst Bogdan et al. (2020) are *rather reserved with respect to the proposed algorithm*, we believe that most of their concerns are actually based on misunderstandings of the algorithm and we are glad to have the opportunity to clarify some of these points. The question of correlated regressors has been addressed in Section 2.1 of this rejoinder, where we have seen that GMJMCMC works reasonably well even when regressors are heavily dependent. Furthermore, Bogdan et al. (2020) were wondering about a *lack of treatment for tautologies*. This can be easily addressed because, in fact, our implementation of GMJMCMC is taking care of Boolean equivalence already when generating new trees. In particular, as we discuss in Section 2.3 of the paper, "*for all three operators it holds that if the newly generated tree is already present in $\mathcal{S}_t$ then it is not considered for $\mathcal{S}_{t+1}$ but rather a new replacement tree is proposed instead.*" What we do in practice is to check whether newly generated trees have correlation $\pm 1$ with any tree within $\mathcal{S}_t$, which for sufficiently large sample size will correspond to logic equivalence. Consequently, tautologies within a GMJMCMC chain are simply not allowed.

Bogdan et al. (2020) also wonder why *the sum in the denominator of (0.2) contains only $M_{fin} = 10000$ models based on d trees from the final stage of the algorithm*. This is indeed one of the implemented options in our package (though $M_{fin}$ does not have to be 10000). The reason for this choice is to avoid having either too large and/or too densely filled hash tables (as a data structure), both of which become quite slow to handle. Whilst this introduces some undesired limitations, it remains an important pragmatic decision to make. The number of logic trees grows exponentially with the number of leaves involved and the number of models grows exponentially with the number of logic trees. Hence, even for the small examples with $p = 50$, the size of a hash table including all visited models and their statistics can become prohibitively large to be used in practice. That would be even more acute for larger $p$'s. As an alternative, one could use the best $N_H$ models from all $T$ generations, where $N_H$ is finite and of reasonable size. But in this case, when the hash table is filled, the worst models must be deleted to allow new ones to be included. In practice, this strategy would become extremely slow. One has to read from, write to and delete from the almost full hash table, which will be also very large. One would either have to create some novel hashing/dehashing functions which make this approach efficient or devise an alternative data structure which is especially designed for the problem at hand. Given the complexity of enumerating logic expressions due to logic equivalence and due to the super-exponential growth of the number of models with respect to the number of leaves involved. We would expect this to be a ground breaking task in the field of algorithms and data structures.

Bogdan et al. (2020) raised the question of why we need MJMCMC for the final population of GMJMCMC when for $d = 15$ full enumeration is feasible. The simple answer is that for many applications one needs much larger $d$ to obtain reliable results, see for example the remark after Theorem 1 of Hubin et al. (2020a) and also Figure 1, panel 3, from the sensitivity analysis of Hubin et al. (2020a). For larger $d$, a full enumeration will no longer be possible, whilst we would like to offer a generally functioning algorithm. Bogdan et al. (2020) additionally say: *"Also, a huge random reduction of the final model space leads to substantially different results for different parallel runs of the algorithm. Therefore the authors aggregate results from different runs using a weighting scheme specified in equation (15) of their paper. In our opinion it seems more reasonable to estimate the posterior probabilities of different models simply by including all models visited in different runs in denominator of (0.2)."* We agree that in principle this is a reasonable approach, which we, in fact, suggested in Section 2.3 of our paper. There, however, we also discussed the drawback that this approach is computationally more costly because one has to transfer a large amount of information from different models between the cores. Finally, Bogdan et al. (2020) promote using synchronization between the cores via priority queues. Whilst we find the idea interesting, we are a little bit sceptical whether it would actually work. When compared to embarrassing parallelization, synchronization between the processes in practice often slows down the inference instead of speeding it up (Chai and Bose, 1993; Kukanov, 2008). There, of course, a lot depends on the back-end used for implementation. We currently do not have the capacity to try this approach ourselves, but we would like, by all means, to encourage Bogdan et al. (2020) or other future researchers of BLR to test this idea. We would be very happy if using synchronization via priority queues could lead to an objectively better and faster inference algorithm for BLR than GMJMCMC.

### 4.1 Mode jumping Markov Chain Monte Carlo

Both Bogdan et al. (2020) and Clarte and Robert (2020) seem to be slightly confused with respect to the MJMCMC algorithm (Hubin and Storvik, 2018), which we did not describe in detail in Hubin et al. (2020a). We thus briefly discuss the main ideas of MJMCMC to clarify certain misunderstandings.

In Hubin and Storvik (2018), a proper MCMC algorithm for the search through a fixed limited model space was proposed. The algorithm deals with the multimodality in the space of models through mode jumping proposals. The mode jumping MCMC (MJMCMC) algorithm relies upon the idea of making smart moves between local extrema with a reasonable frequency. Local MCMC is performed in the absolute majority of steps. For the rest, a large move in the model space (which is likely to hit a model with very low posterior probability) is made, followed by local optimization. The goal of the latter step is to reach a local optimum in a different part of the model space. Then the proposal is randomized around this optimum and the transition to the proposed model is either accepted or rejected according to a Metropolis-Hastings acceptance probability. The convergence properties of the suggested Markov chain is proven through a refinement of the results of Tjelmeland and Hegstad (2001). Its limiting distribution is shown to correspond to the marginal model posterior probabilities. Further extensions of the algorithm allowing for parallel computing and using mixtures of proposals were also suggested.

MJMCMC is described in more detail in Algorithm 1 below, where we consider $M = (\gamma_1, ...\gamma_p)$ to be associated with models in the given discrete model space $\Omega$ (here $\gamma_j \in \{0, 1\}$ indicates whether covariate $x_j$ is included in the model). We assume that marginal likelihoods $p(Y|M)$ are available for a given $M$, and then use MJMCMC to explore $p(M|Y)$. By Bayes formula

$$p(M|Y) = \frac{p(Y|M)p(M)}{\sum_{M' \in \Omega} p(Y|M')p(M')}. \qquad (2)$$

In order to calculate $p(M|Y)$ we have to iterate through the whole model space $\Omega$, which becomes computationally infeasible for large $p$. The ordinary Monte Carlo estimate is based on a number of MJMCMC samples $M^{(i)}, i = 1, ..., W$:

$$\widetilde{p}(M|Y) = \frac{1}{W} \sum_{i=1}^{W} \mathbb{1}(M^{(i)} = M) \xrightarrow[W \to \infty]{d} p(M|Y), \qquad (3)$$

where $\mathbb{1}(\cdot)$ is the indicator function. An alternative named the renormalized model (RM) estimate by Clyde et al. (2011), is

$$\widehat{p}(M|Y) = \frac{p(Y|M)p(M)}{\sum_{M' \in \mathbb{V}} p(Y|M')p(M')} \mathbb{1}(M \in \mathbb{V}), \qquad (4)$$

where now $\mathbb{V}$ is the set of **all models visited at least once** during the MJMCMC run. Assuming the Markov chain eventually will visit all possible models, also $\widehat{p}(M|Y)$ will converge to $p(M|Y)$. Note that this estimate also can utilize all models that are visited, not only those that have been accepted. This answers the comment of Bogdan et al. (2020), who presumed that we include only models accepted by MJMCMC into $\mathbb{V}$. Although both (4) and (3) are asymptotically consistent, (4) will often be the preferable estimator since the convergence of the MCMC based approximation (3) is typically much slower, see Clyde et al. (2011).

We now describe the MJMCMC algorithm in more detail. We aim at approximating $p(M|Y)$ by means of searching for some subspace $\mathbb{V}$ of $\Omega$ which makes the approximation (4) as precise as possible. Models with high values of $p(Y|M)$ are important to be included. This means that modes and near modal values of marginal likelihoods are particularly important for the construction of $\mathbb{V} \subset \Omega$ and missing them can dramatically influence our estimates. Note that these considerations are equally important for the standard MCMC estimate (3). The main difference is that when using (3) the number of times a specific model is visited is important, for (4) it is enough that a model is visited at least once. In this context, the denominator of (4) becomes an extremely relevant measure for the quality of the search. It should be as large as possible in order to capture the probability mass from all the local optima of the posterior distribution, whilst at the same time the size of $\mathbb{V}$ should be low in order to save computational time.

---

**Algorithm 1** Mode jumping MCMC

---

1: Generate a large jump $M_0^*$ according to a proposal distribution $q_l(M_0^*|M)$.
2: Perform a local optimization, defined through $M_k^* \sim q_o(M_k^*|M_0^*)$.
3: Perform a small randomization to generate the proposal $M^* \sim q_r(M^*|M_k^*)$.
4: Generate backwards auxiliary variables $M_0 \sim q_l(M_0|M^*)$, $M_k \sim q_o(M_k|M_0)$.
5: Put
$$M' = \begin{cases} M^* & \text{with probability } r_{mh}(M, M^*; M_k, M_k^*); \\ M & \text{otherwise,} \end{cases}$$
where
$$r_{mh}^*(M, M^*; M_k, M_k^*) = \min\left\{1, \frac{p(M^*|y)q_r(M|M_k)}{p(M|y)q_r(M^*|M_k^*)}\right\}. \tag{5}$$

---

Algorithm 1 describes in detail the mode jumping step within the MJMCMC algorithm. In the first step, a large change in the model space is made through the proposal distribution $q_l$. This will typically lead to a model with little support in the data, so in step 2 a local optimization is performed in order to obtain a better model. Due to the need for a proper Metropolis-Hastings probability derived through a backwards move (step 4), a randomization, through $q_r$, of the local optima is needed for the reverse move back to the original model to be possible. Step 5 specifies the acceptance probability which is shown in Hubin and Storvik (2018) to satisfy the detailed balance equation with respect to $p(M|Y)$.

Hopefully, this detailed discussion of MJMCMC fully resolves the confusion of Clarte and Robert (2020), who, in their discussion, presume the following: *"While we have not read the refered article on MJMCMC in detail, a first comment is that the name itself is somewhat unsuitable, as indeed the algorithm does not sample from a distribution but only explores its surface."* We would like to emphasize that the MJMCMC ***is not*** incorporating any of the ideas of Tabu search algorithms (Glover et al., 1995), which are not allowing to return to the previously visited models. This should also clarify another misleading presumption by Clarte and Robert (2020): *"In a self-avoiding mode, keeping track of all the previous states visited by the chain ensures that those states will never be visited again. As we are in a discrete setting, this implies that once a mode has been visited the algorithm is constrained to eventually visit another mode, even if the potential between the modes is almost zero."*

**Convergence of GMJMCMC**

The MJMCMC algorithm, in the setting of BLR, only gives convergence within each of the restricted search spaces (populations) that it considers. We apply the MJMCMC as an inner iteration within the GMJMCMC algorithm where the space of models is dynamically modified. Given that the movement within and between the search spaces is irreducible with respect to the whole model space, which is shown in Theorem 1 of Hubin et al. (2020a), the GMJMCMC provides the estimates equivalent to (4). They also converge towards the right model probabilities. This fully resolves another concern from Clarte and Robert (2020) who stated that the renormalized estimator of the marginal posterior model probabilities *"does not provide the theoretical security of (asymptotic) unbiasedness that is attained with MCMC method."*

## 4.2 Parameter settings

The choice of the tuning parameters for the algorithm is definitely an important problem as indicated by Ruczinski et al. (2020) and Clarte and Robert (2020). Whilst there is not (and cannot be) any uniformly best choice of tuning parameters of GMJMCMC, we will try to briefly indicate some strategies allowing to manually choose reasonable values of the most important tuning parameters of the algorithm. Regarding the choice of the population size $d$ and the maximal number of variables in a model $k_{max}$, we give some guidance in Remark 1 after Theorem 1 in Hubin et al. (2020a): *"When $d_1 > 0$ (which is the $N_{init}$ covariates with largest marginal inclusion probability in $\mathcal{S}_1$), some restrictions on the possible search spaces are introduced. However, when $d - d_1 \geq k_{max}$, any model of maximum size $k_{max}$ will eventually be visited. If $d - d_1 < k_{max}$, then every model of size up to $d - d_1$ plus some of the larger models will eventually be visited, although the model space will get some additional constraints. In practice, it is more important that $d - d_1 \geq k^*$, where $k^*$ is the size of the true model. Unfortunately, neither $k^*$ nor $d_1$ are known in advance, and one has to make reasonable choices of $k_{max}$ and $d$ depending on the problem one analyses."* Also, note that we provide some sensitivity analysis of $d$ in Section 3.1 of the main article.

Regarding the maximal depth of logic expressions $C_{max}$, one should use some prior knowledge on the complexity of logic expressions. It also depends upon the individual hypotheses the researcher has. At the same time, using unreasonably large $C_{max}$ is prohibitive computationally and also unrealistic in terms of power to detect too complex trees.

When combining two Boolean expressions, first a decision is made whether it will be combined through an *and* or an *or* operation (with $P_{and}$ specifying the probability for *and*) and thereafter a decision is made whether the logic *not* is applied to it (with probability $P_{not}$). In our experience, the actual values of these tuning parameters will not influence the result very much with respect to finding the right expressions within the equivalence classes. However, simpler expressions (within the equivalence classes) are usually obtained when choosing somewhat larger $P_{and}$ and somewhat smaller $P_{not}$. We recommend the choice $P_{and} = 0.9$ and $P_{not} = 0.1$.

The tuning parameter $\rho_{min}$ is used to determine which variables should be removed from the current population with probability one minus the current approximation of the marginal inclusion probability of these variables. $\rho_{min}$ should be chosen in such a way that it is on the one hand possible to get rid of unimportant trees, while at the same time avoiding the deletion of potentially important trees. Concerning the question of Ruczinski et al. (2020) on the choice of $M_{fin}$ and $T_{max}$ and the resulting chain length, we provided some guidance in Theorem A.1 in Section A.2 of Hubin et al. (2020b). There, we proved convergence guarantees also for fixed $T_{max}$ and $M_{fin}$ when increasing the number of parallel chains of GMJMCMC. Thus, apparently, there exists a natural trade off: the more chains one can afford running in parallel the fewer resources could be used within each chain and vice versa - the less parallel chains one runs - the larger $T_{max}$ and $M_{fin}$ are required.

The choice of the tuning parameters for the examples from Hubin et al. (2020a) are provided in Section A.1 of Hubin et al. (2020b). These values might be considered for problems of similar dimensionality, effect sizes and correlations between covariates. At the same time, we cannot provide a strict stopping criterion for GMJMCMC or a general rule for the choice of its parameters. Experimental tuning for different applications might be beneficial. If one has enough computational resources, grid search or an adaptation of Bayesian optimization for the tuning parameters of GMJMCMC (Snoek et al., 2012) can be considered. Alternatively, one might consider some kind of adaptive learning of the algorithm's tuning parameters similarly to Hubin (2019). More details on these possibilities are beyond the scope of this rejoinder.

## 5 Various extensions of BLR and GMJMCMC

In this section, we briefly present extensions of the logic regression model. Some of these extensions are further discussed in the tutorial of Section 6 of the rejoinder. A more detailed description of the proposed extensions, including theoretical support and real applications, are material for a future publication.

### 5.1 Predictions with BLR

As mentioned in the discussion section of Hubin et al. (2020a), our method is directly applicable to prediction as well. In particular, the standard Bayesian model averaging can be easily applied. Thus, one can approximate the posterior probability of some parameter/variable $\Delta$ via model averaging by

$$\hat{p}(\Delta \mid Y) = \sum_{M \in \mathbb{V}} p(\Delta \mid M, Y)\hat{p}(M \mid Y) , \qquad (6)$$

where $\Delta$ might be, for example, the predictor of unobserved data based on a specific set of covariates. Given estimates of model posterior probabilities, other prediction procedures such as the median probability model (Barbieri and Berger, 2004) or the most probable model can be also easily adopted, yielding:

$$\hat{p}(\Delta \mid Y) = p(\Delta \mid M^*, Y) , \qquad (7)$$

where $M^*$ is the selected median probability or the most probable a posteriori model.

## 5.2 BLR with non-binary covariates

Responding to a question from Ruczinski et al. (2020), we can allow non-binary fixed effects to be included in the model. For this extension, we simply replace equation (2) in Hubin et al. (2020a) with:

$$\mathsf{h}\left(\mu\left(\boldsymbol{X}\right)\right) = \alpha + \sum_{j=1}^{q} \gamma_j \beta_j L_j + \sum_{j=q+1}^{q+q'} \gamma_j \beta_j z_{j-q}, \tag{8}$$

where $z_l, l \in \{1, ..., q'\}$ are non-binary covariates which are not allowed to form logic expressions. In this formulation of the Bayesian logic regression, the model includes $q+q'$ possible components. The priors on the additional components $\gamma_j, j \in \{q+1, ..., q+q'\}$ are of form (4) from Hubin et al. (2020a) with $c(z_{j-q}) = 1, j \in \{q+1, ..., q+q'\}$. This results in the following joint model prior:

$$p(M) \propto a^{\sum_{j'=q+1}^{q+q'} \gamma_{j'}} \prod_{j=1}^{q} a^{\gamma_j c(L_j)}. \tag{9}$$

In terms of model inference, the GMJMCMC is adopted, where modifications, mutations and reductions are only allowed for the Boolean terms.

## 5.3 BLR with non-binary covariates and latent Gaussian variables

We also mentioned in Hubin et al. (2020a) that it is straight-forward to extend our approach for generalized linear mixed models. Here, we will formally describe this extension by including both fixed effects for non-binary covariates and latent Gaussian variables, which can be used to model correlation structures between observations (in space and time) and over-dispersion. For this extension, we further update equation (8),

$$\mathsf{h}\left(\mu\left(\boldsymbol{X}\right)\right) = \alpha + \sum_{j=1}^{q} \gamma_j \beta_j L_j + \sum_{j=q+1}^{q+q'} \gamma_j \beta_j z_{j-q} + \sum_{k=1}^{r} \delta_{ik}, \tag{10}$$

where $z_l, l \in \{1, ..., q'\}$ are non-binary covariates which are not allowed to form logic expressions and $\boldsymbol{\delta_k} = (\delta_{1k}, ..., \delta_{nk}) \sim N_n\left(\boldsymbol{0}, \boldsymbol{\Sigma}_k\right)$ are latent Gaussian variables. The latent Gaussian variables with covariance matrices $\boldsymbol{\Sigma}_k$ allow to model different correlation structures between individual observations (e.g. auto-regressive models or various other spatio-temporal models). The matrices typically depend only on a few parameters, so that in practice one has $\boldsymbol{\Sigma}_k = \boldsymbol{\Sigma}_k(\boldsymbol{\psi}_k)$. Whilst the model priors (9) are still valid, parameter priors here need to be adjusted as

$$\boldsymbol{\beta}|\boldsymbol{\gamma} \sim N_{p_{\boldsymbol{\gamma}}}(\boldsymbol{0}, I_{p_{\boldsymbol{\gamma}}} e^{-\psi_{\beta_{\boldsymbol{\gamma}}}}), \tag{11}$$

$$\boldsymbol{\psi}_k \sim \pi_k(\boldsymbol{\psi}_k). \tag{12}$$

Here, all kind of hyper-parameters of priors compatible with INLA (Rue et al., 2009) can in principle be chosen. This allows to efficiently compute the marginal likelihoods of individual models using the INLA approach (Rue et al., 2009; Hubin and Storvik, 2016).

# 6  A tutorial on GMJMCMC for BLR

Finally, we provide a brief tutorial on how to apply our approach in practice. Our code should be run under Linux. One would need to incorporate some sort of extra *hacks* (see https://bit.ly/37tf3cm) to be able to run the code under Windows (due to the limitations of the standard ***parallel::mclapply*** R function which is applied within the library).

## 6.1  Installing the packages

We start by preparing the R environment for running our approach to BLR. The R-script below will install all packages that are needed to run the code. Depending on which R packages you have already installed, running this script might take a while. Then we install the EMJMCMC package from GitHub.

```r
#**********************************************************************
# install all packages which will be needed for the EMJMCMC package
source("https://raw.githubusercontent.com/aliaksah/EMJMCMC2016/master/
R/load_dependencies/loaddeps.R")
#**********************************************************************
# (currently works only under Linux)
install.packages("https://github.com/aliaksah/EMJMCMC2016/blob/master/
EMJMCMC_1.4.2_R_x86_64-pc-linux-gnu.tar.gz?raw=true",
repos = NULL, type="source")
#**********************************************************************
```

One might want to restart R before proceeding to have a clean environment. After having the package installed we can load EMJMCMC.

```r
#**********************************************************************
# load the EMJMCMC package
library(EMJMCMC)
#**********************************************************************
```

Additionally, we will need the following three packages for the tutorial, which you might have to install from CRAN.

```r
#*********************************************************************
# load other packages needed to simulate and illustrate data
# if necessary these packages first have to be installed from CRAN
library(clusterGeneration)
library(bindata)
library(ggplot2)
#*********************************************************************
```

## 6.2 Running BLR with weakly correlated covariates

We first generate some binary data with the general correlation structure from the first scenario of the simulation study above.

```r
#*********************************************************************
# set the seed
set.seed(040590)
# construct a correlation matrix for M = 50 variables
M = 50
m = clusterGeneration::rcorrmatrix(M,alphad=2.5)
# simulate 1000 binary variables with this correlation matrix
X = bindata::rmvbin(1000, margprob = rep(0.5,M), bincorr = m)
#*********************************************************************
```

The following code generates the heat-map of Figure 1 which illustrates the non-trivial correlations of the simulated binary variables.

```r
#*********************************************************************
# prepare the correlation matrix in the melted format
melted_cormat = reshape2::melt(cor(X))
# plot the heat-map of the correlations
ggplot2::ggplot(data = melted_cormat,
ggplot2::aes(x=Var1, y=Var2, fill=value)) +
  ggplot2::geom_tile() +
  ggplot2::theme(axis.title.x = ggplot2::element_blank(),
                 axis.title.y =  ggplot2::element_blank(),
                 axis.text.x = ggplot2::element_blank())
#*********************************************************************
```

Next, we simulate the responses according to Scenario 4 from Hubin et al. (2020a), but with correlated binary covariates.

```r
1  #**********************************************************************
2  # simulate Gaussian responses from a model with two-way interactions
3  # which is considered in S.4 of the paper
4  df = data.frame(X)
5  df$Y=rnorm(n = 1000,mean = 1+1.43*(df$X5*df$X9)+
6      0.89*(df$X8*df$X11)+0.7*(df$X1*df$X4),sd = 1)
7  #**********************************************************************
```

Before performing logic regression with GMJMCMC one might like to have a look at the documentation of the R function ***LogicRegr***:

```r
1  #**********************************************************************
2  help("LogicRegr")
3  #**********************************************************************
```

The following code runs inference on BLR with 32 parallel threads of GMJMCMC, where we are first using the robust g-prior and then Jeffreys prior. Depending on the cluster each of these might run for some time from several minutes to more than half an hour. If you are running the code on a home PC or a laptop, please reduce *ncores* parameter to something reasonable for your machine (e.g. set *ncores* = 3).

```r
1  #**********************************************************************
2  # specify the initial formula
3  formula1 = as.formula(paste(colnames(df)[M+1],"~ 1 + ",
4      paste0(colnames(df)[-c(M+1)],collapse = "+")))
5  #**********************************************************************
6  # Bayesian logic regression with the robust-g-prior
7  res4G = LogicRegr(formula = formula1, data = df,
8      family = "Gaussian", prior = "G", report.level = 0.5,
9      d = 15,cmax = 2,kmax = 15, p.and = 0.9, p.not = 0.1, p.surv = 0.2,
10     ncores = 32)
11 #**********************************************************************
12 # Bayesian logic regression with the Jeffreys prior
13 res4J = LogicRegr(formula = formula1, data = df,
14     family = "Gaussian", prior = "J", report.level = 0.5,
15     d = 15, cmax = 2,kmax = 15, p.and = 0.9, p.not = 0.1, p.surv = 0.2,
16     ncores = 32)
17 #**********************************************************************
```

We obtain the following results using the robust g-prior:

```
#**********************************************************************
# print the results for the robust g-prior
print(base::rbind(c("expressions","probabilities"),res4G$feat.stat))
     [,1]                 [,2]
[1,] "expressions"        "probabilities"
[2,] "I(((X5))&((X9)))"   "1"
[3,] "I(((X1))&((X4)))"   "1"
[4,] "I(((X11))&((X8)))"  "0.999999645314492"
#**********************************************************************
```

and rather similar results with the Jeffreys prior:

```
#**********************************************************************
#print the results for the Jeffreys prior
print(base::rbind(c("expressions","probabilities"),res4J$feat.stat))
     [,1]                 [,2]
[1,] "expressions"        "probabilities"
[2,] "I(((X11))&((X8)))"  "0.999999774980675"
[3,] "I(((X1))&((X4)))"   "0.999999520871822"
[4,] "I(((X5))&((X9)))"   "0.999873046960372"
#**********************************************************************
```

### 6.3 Additional non-binary fixed effects and predictions

Ruczinski et al. (2020) asked whether it would be possible to include covariates in the model which are not a part of the logic expressions. Furthermore, Schwender and Ickstadt (2020) are interested in whether the model can be easily used for predictions. These options are currently not implemented in the **LogicRegr** function, which we would like to keep as simple as possible. At the same time, these tasks can be easily performed by a general call of the **EMJMCMC::pinferunemjmcmc** function which is available in our package. This routine is however much more advanced and requires, at this time, expert knowledge to be used.

First, we will generate an additional Poisson distributed covariate *age* which is then used as an additional additive effect in the data generating logic regression model. For the sake of brevity we perform the analysis here only with Jeffreys prior.

```
#**********************************************************************
# simulate Gaussian responses from a model with two-way interactions
# and an age effect which is an extension of S.4 of the paper
Xp = data.frame(X)
```

```
5   Xp$age = rpois(1000,lambda = 34)
6   Xp$Y=rnorm(n = 1000,mean = 1+0.7*(Xp$X1*Xp$X4) +
7   0.89*(Xp$X8*Xp$X11)+1.43*(Xp$X5*Xp$X9) + 2*Xp$age, sd = 1)
8   #**********************************************************************
```

We will not only perform model inference but also show how to make predictions with the *EMJMCMC* package. To this end we will randomly divide the data into a training set (900 observations) and a testing set (100 observations).

```
1   #**********************************************************************
2   teid  = sample.int(size =100,n = 1000,replace = F)
3   test  = Xp[teid,]
4   train = Xp[-teid,]
5   #**********************************************************************
```

The function **pinferunemjmcmc** has more capabilities than performing logic regression. First, one might want to see its arguments:

```
1   #**********************************************************************
2   help("pinferunemjmcmc")
3   #**********************************************************************
```

The following call of **pinferunemjmcmc** performs logic regression using 30 cores. Note that the non-binary covariate *is not* a part of the formula passed to the function, but is rather specified through *runemjmcmc.params*$*latnames* = "$I(age)$". Also, one might expect this to run slightly longer than previous examples, particularly because keeping track of the $\beta$ coefficients for prediction takes some additional time. Further, many of the input options used are explained in the help pages of **pinferunemjmcmc**. If one is not interested in predictions, *runemjmcmc.params*$*save.beta* = $F$, *predict* = $F$ and *test.data* = $NULL$ should be set (this will decrease inference time for the same training data sample and other tuning parameters fixed).

```
1   #**********************************************************************
2   # specify the link function
3   g = function(x) x
4   #**********************************************************************
5   # specify the parameters of the custom estimator function
6   estimator.args = list(data = train, n = dim(train)[1],
7     m =stri_count_fixed(as.character(formula1)[3],"+"),k.max = 15)
8   #**********************************************************************
```

```
9    # specify the parameters of gmjmcmc algorithm
10   gmjmcmc.params = list(allow_offsprings=1,mutation_rate = 250,
11     last.mutation=10000, max.tree.size = 5, Nvars.max =15,
12     p.allow.replace=0.9,p.allow.tree=0.01,p.nor=0.01,p.and = 0.9)
13   #*********************************************************************
14   # specify some advenced parameters of mjmcmc
15   mjmcmc.params = list(max.N.glob=10, min.N.glob=5, max.N=3, min.N=1,
16     printable = F)
17   #*********************************************************************
18   # run the inference of BLR with a non-binary covariate and predicions
19   res.alt = pinferunemjmcmc(n.cores = 30, report.level =  0.2,
20     num.mod.best = 100,simplify = T,predict = T,test.data = test,
21     link.function = g,
22     runemjmcmc.params = list(formula = formula1,latnames = c("I(age)"),
23      data = train,estimator = estimate.logic.lm.jef,
24      estimator.args =estimator.args,
25      recalc_margin = 249, save.beta = T,interact = T,outgraphs=F,
26      interact.param = gmjmcmc.params,
27      n.models = 10000,unique = T,max.cpu = 4,max.cpu.glob = 4,
28      create.table = F,create.hash = T,pseudo.paral = T,burn.in = 100,
29      print.freq = 1000,
30      advanced.param = mjmcmc.params))
31   #*********************************************************************
```

Below a list of the logic expressions and non-logic covariates that were found to be of importance is listed. There, we clearly see that all features from the data-generative model are detected without any false positive discoveries.

```
1   #*********************************************************************
2   print(base::rbind(c("expressions","probabilities"),res.alt$feat.stat))
3        [,1]                [,2]
4   [1,] "expressions"       "probabilities"
5   [2,] "I(((X5))&((X9)))"  "1"
6   [3,] "I(age)"            "0.999999999999998"
7   [4,] "I(((X11))&((X8)))" "0.999999990458405"
8   [5,] "I(((X1))&((X4)))"  "0.99999997999928"
9   #*********************************************************************
```

To assess the quality of prediction we use two criteria, RMSE = $\sqrt{n_p^{-1} \sum_{i=1}^{n_p}(\hat{Y}_i^* - Y_i^*)^2}$ and MAE = $n_p^{-1} \sum_{i=1}^{n_p} |\hat{Y}_i^* - Y_i^*|$, where $Y_i^*$ are responses in the test data, $\hat{Y}_i^*$ are model averaged predictions of them, and $n_p$ is the size of the test data set.

```
1  #************************************************************************
2  print(sqrt(mean((res.alt$predictions-test$Y)^2)))
3  [1] "0.8835489"
4  print(mean(abs((res.alt$predictions-test$Y))))
5  [1] "0.6904736"
6  #************************************************************************
```

We want to compare the performance of BLR in this example with a simple standard approach, namely ridge regression (Zou and Hastie, 2005), combined with model selection according to AIC. In the script below we run ridge regression and perform prediction on the test data set.

```
1   #************************************************************************
2   library(HDeconometrics)
3   ridge = ic.glmnet(x = train[,-51],y=train$Y,family = "gaussian",
4   alpha = 0)
5   predict.ridge = predict(ridge$glmnet,newx = as.matrix(test[,-51]),
6   type = "response")[,which(ridge$glmnet$lambda == ridge$lambda)]
7   print(sqrt(mean((predict.ridge-test$Y)^2)))
8   [1] "1.061406"
9   print(mean(abs((predict.ridge-test$Y))))
10  [1] "0.865467"
11  #************************************************************************
```

We finally compute the evaluation metrics for prediction based on the expectations of the data-generative (true) model for the test data:

```
1  #************************************************************************
2  tmean = 1+2*test$age+0.7*(test$X1*test$X4) +
3  0.89*(test$X8*test$X11)+1.43*(test$X5*test$X9)
4  print(sqrt(mean((tmean - test$Y)^2)))
5  [1] "0.8671786"
6  print(mean(abs((tmean - test$Y))))
7  [1] "0.6850737"
8  #************************************************************************
```

We clearly see that for this specific example logic regression significantly outperforms the ridge regression baseline with respect to both RMSE and MAE. This is not surprising given that the data generative process has multiple non-linear effects. Moreover, the predictions obtained by the BLR model are extremely close to the predictions from the means of the data generative model.

# 7 Comparison with other approaches

Several other approaches were mentioned by the discussants. Ruczinski et al. (2020) mentioned that simulated annealing for logic regression could be equipped with a penalized likelihood criterion following from the priors used in our setting. Schwender and Ickstadt (2020) pointed out certain similarities of GMJMCMC with Genetic Programming for Association Studies as well as logic Feature Selection. Bogdan et al. (2020) mentioned the recently developed non-reversible MCMC algorithms as well as parallel tempering MCMC algorithms. It would be most interesting to compare all these different algorithms with GMJMCMC but we believe this would need substantial additional effort and goes far beyond the scope of this rejoinder. We leave these possibilities open as topics for further research.

# 8 Conclusions

We would like to thank once again all of the discussants for their valuable and insightful feedback. We are happy to have provoked so many questions, comments and remarks. We hope that we managed to shed light on the majority of them in this rejoinder. Moreover, we provided some useful extension of Bayesian logic regression method here. The discussions also motivate multiple directions for further research, which are outside the scope of this rejoinder. However, we hope this research will be in future performed in close collaboration with the discussants.